\documentclass[11pt]{article}
\newcommand{\D}{\displaystyle}
\newcommand{\ltbar}{\left|\!\left|\!\left|}
\newcommand{\rtbar}{\right|\!\right|\!\right|}
\newcommand{\mbf}[1]{\mbox{\boldmath{$#1 $}}}

\begin{document}

\title{Program to calculate pure angular momentum coefficients
       in $jj$--coupling}

\author{ Gediminas Gaigalas$^{\, a,b}$, Stephan Fritzsche$^{a}$ and
         Ian P. Grant$^{c}$
         \\
         \\
         \\
        $^a$ Fachbereich Physik, Universit\"a{}t Kassel,  \\
        Heinrich--Plett--Str. 40, D--34132 Kassel, Germany.
        \\
        $^b$ Institute of Theoretical Physics and Astronomy,  \\
        A.\ Go\v{s}tauto 12, Vilnius 2600, Lithuania.
        \\
        $^c$ Mathematical Institute, University of Oxford,   \\
        24/29 St. Giles', Oxford OX1 3LB, UK.
        \\}

\maketitle

\date{}

\thispagestyle{empty}

\begin{abstract}
A program for computing pure angular momentum coefficients in
relativistic atomic structure for any scalar one-- and
two--particle operator is presented.  The program, written in
Fortran 90/95 and based on techniques of second quantization,
irreducible tensorial operators, quasispin and the theory of
angular
momentum, is intended to replace existing angular coefficient modules
from GRASP92. The new module uses a different decomposition of
 the
coefficients as sums of products of pure angular momentum coefficients,
 which
depend only on the tensor rank of the interaction but not on its
details, with effective interaction strengths of specific
interactions.  This saves memory and reduces the computational cost of
big calculations signficantly. 
\end{abstract}

\newpage

{\large\bf PROGRAM SUMMARY}

\bigskip

{\it Title of program:} ANCO

\bigskip

{\it Catalogue number:} ADOO

\bigskip

{\it Program obtainable from:} CPC Program Library,
     Queen's University of Belfast, N. Ireland. Users may obtain the
     program also by down--loading the tar--file \texttt{ratip-anco.tar}
     from our home page at the University of Kassel    \newline
 (http://www.physik.uni-kassel.de/fritzsche/programs.html).

\bigskip

{\it Program Summary URL:} http://cpc.cs.qub.ac.uk/summaries/ADOO

\bigskip

{\it Licensing provisions:} None.

\bigskip

{\it Computer for which the program is designed and has been tested:}
 \newline
     IBM RS 6000, PC Pentium II.                            \newline
     {\it Installations:} University of Kassel (Germany).   \newline
     {\it Operating systems:} IBM AIX 4.1.2+, Linux 6.1+.

\bigskip

{\it Program language used in the new version:} ANSI standard Fortran
 90/95.

\bigskip

{\it Memory required to execute with typical data:} 100 kB.

\bigskip

{\it No.\ of bits in a word:}  All real variables are parametrized by a
     \texttt{selected kind parameter}. Currently this is set to double
     precision for consistency with other components of the RATIP
     package [1]. 
\bigskip

{\it Distribution format:} Compressed tar file. On a UNIX (or compatible)
     workstation, the command \texttt{uncompress} uncompress this file and
     the command \texttt{tar -xvf ratip\_anco.tar} reconstructs the file
     structure.

\bigskip

{\it Keywords:} atomic many--body perturbation theory, complex atom,
     configuration interaction, effective Hamiltonian, energy level,
     Racah algebra, reduced coefficients of fractional parentage,
     reduced matrix element, relativistic, second quantization,
     standard unit tensors, tensor operators,
     $9/2$--subshell.

\bigskip

{\it Nature of the physical problem:}  \newline

     The matrix elements of a one--electron tensor operator $\widehat{A}^k$
     of
     rank $k$ with respect to a set of configuration state functions
     $|\gamma_iJ_i\rangle$ can be written $\sum_{ab} t^{k}_{ij}(ab)\,
     (a|\widehat{A}^k |b)$ where the angular coefficients $t^{k}_{ij}(ab)$
     are independent of the operator $\widehat{A}^k$, $i,j$ are
     CSF labels and $a,b$ run over the relevant interacting orbital
     labels. Similarly, the matrix elements of the Dirac--Coulomb
     Hamiltonian can be written in the form $\sum_{ab} t^{0}_{ij}(ab)\,
     (a|\widehat{H}_D |b) +\sum_k \sum_{abcd} v^k_{ij}(abcd)\,
     X^k(abcd)$, where $\widehat{H}_D$ is the one-electron Dirac
     Hamiltonian operator, with tensor rank zero, $v^k_{ij}(abcd)$
     are pure angular momentum coefficients for two--electron
     interactions, and $X^k(abcd)$ denotes an effective interaction
     strength for the two electron interaction. The effective
     interaction strengths for Coulomb and Breit interaction have
     different selection rules and make use of subsets of the full set
     of coefficients $v^k_{ij}(abcd)$.
   
     Such
     matrix elements are required for the theoretical determination of
     atomic energy levels, orbitals and radiative transition data in
     relativistic atomic structure theory.  The code is intended for
     use in configuration interaction or multiconfiguration Dirac--Fock
     calculations~[2], or for calculation of matrix elements of the
     effective Hamiltonian in many--body perturbation
     theory~[3].

\bigskip

{\it Method of solution:}  \newline
     A combination of second quantization and quasispin methods
     with the theory of angular momentum and irreducible tensor operators
     leads to a more efficient evaluation of (many--particle) matrix
     elements and to faster execution of the code~[4].

\bigskip

{\it Restrictions on the complexity of the problem:}  \newline
     Tables of reduced matrix elements of the tensor operators
     $a^{(q~j)}$ and $W^{(k_q k_j)}$ are provided for
     ($nj$) with $j = 1/2, 3/2, 5/2, 7/2$, and $9/2$.
     Users wishing to extend the tables must provide the
     necessary data.

\bigskip

{\it Typical running time:}  \newline
     3.5 seconds for both examples on a 450 MHz Pentum III
     processor.

\bigskip

{\it Unusual features of the program:} \newline
     The program is designed for large--scale atomic structure
     calculations and its computational cost is less than that of the
     corresponding angular modules of GRASP92.  The present version of
     the program generates pure angular momentum coefficients
     $t^{0}_{ij}(ab)$ and $v^k_{ij}(abcd)$, but coefficients
     $t^{k}_{ij}(ab)$ with $k>0$ have not been enabled.  An option is
     provided for generating coefficients compatible with existing
     GRASP92.

     Configurational states with any distribution of electrons in shells
     with $j \leq 9/2$ are allowed. 
     This pemit user to take into account the single, double, triple excitations form
     open $d$-- and $f$-- shells for the systematic MCDF studies of heavy and
     superheavy elementds (Z $>$ 95).

\bigskip

{\it References:}   \newline
     [1] S.\ Fritzsche, C.\ F.\ Fischer, and C.\ Z.\ Dong,
         Comput.\ Phys.\ Commun.\ 124 (1999) 240.
     \newline
     [2] I.\ P.\ Grant, Methods of Computational Chemistry, Vol
           2. (ed. S.\ Wilson) pp. 1-71 (New York, Plenum Press, 1988);
           K.\ G.\ Dyall, I.\ P.\ Grant, C.\ T.\ Johnson, F.\ A.\
           Parpia and E.\ P.\ Plummer,  Comput.\ Phys.\ Commun.\ 55
           (1989) 425; F.\ A.\ Parpia, C.\ Froese Fischer and I.\ P.\
           Grant, Comput.\ Phys.\ Commun.\ 92 (1996) 249.
     \newline
     [3] G.\ Merkelis, G.\ Gaigalas, J.\ Kaniauskas, and Z.\ Rudzikas,
         Izvest.\ Acad.\ Nauk SSSR, \newline
     \hspace*{0.55cm}Phys.\ Series 50 (1986) 1403.
     \newline
     [4] G.\ Gaigalas, Lithuanian Journal of Physics 39 (1999)
     80.

\newpage
{\large\bf LONG WRITE--UP}

\bigskip

\section{Introduction}

The improved accuracy of modern experiments challenges theorists to
match or
exceed experimental precision. Models of many--electron atoms and
ions require
both relativistic and correlation effects to be taken into account;
this can
be done, for example, by  using various versions of perturbation
theory,
the configuration interaction method, the  multiconfiguration
Hartree--Fock method
\cite{kn:fa} or the multiconfiguration Dirac--Fock
method
\cite{kn:gk}.

\medskip

The evaluation of matrices of one-- and two--electron
operators for
many--electron states in $jj$--coupling is customarily done by
expressing each matrix element as a sum of products of angular
coefficients and radial integrals.  This strategy, based on earlier
work by Fano, was adopted for the MCP program~\cite{MCP,MCP75}
for evaluating
angular coefficients for the Coulomb interaction and the
related MCBP
program~\cite{MCBP} for evaluating angular coefficients for
the Breit
interaction.  Whilst this strategy was adequate on the
relatively low--powered
computers of the 1970s, it is now possible and desirable to use very
large configuration sets which require a more efficient strategy.
The new program attains this objective by building up
the angular
coefficients for one-- and two--electron interactions from a
relatively small number of common spin-angular parts in the
manner of~\cite{method2,method6,method7}.

\medskip

The theoretical background is presented in Section~2,
program organization
in section~3, testing and timing studies in section~4 and
simple test problems in Section~5.

\section{Theoretical background}

\subsection{Dirac--Coulomb Hamiltonian}

As usual, multiconfiguration self--consistent--field calculations
in relativistic atomic theory are based on the
Dirac--Coulomb hamiltonian:
\begin{eqnarray}
\label{eq:DC-a}
\D  \widehat{H}_{DC}=
    {\sum_{i = 1}^{N}}
    \widehat{H}_{D}(i) +
    \frac{1}{2} {\sum_{i,j = 1}^{N}}
    \frac{1}{r_{ij}}.
\end{eqnarray}
$\widehat{H}_{D}(i)$ is the Dirac one--particle
operator,
\begin{eqnarray}
\label{eq:DC-b}
\D  \widehat{H}_{D}=
    c {\mbf \alpha}(i) \cdot {\mbf p}_i + [\beta(i) -1] c^2
    -\frac{Z}{r_i},
\end{eqnarray}
and the second term in (\ref{eq:DC-a}) represents the Coulomb
interaction of pairs of electrons. In equation (\ref{eq:DC-b})
$c$ denotes the speed of light; ${\mbf \alpha}(i)$ and
$\beta(i)$ are 4 $\times$ 4 Dirac matrices for the $i$th electron;
$r_i$ and ${\mbf p}_i$ are the radial coordinate of the $i$th electron
and its (3--) momentum, respectively.
The first two terms of (\ref{eq:DC-b}) comprise the Dirac kinetic
energy operator.

\medskip

Both configuration interaction and multiconfiguration
Dirac--Hartree--Fock calculations require the matrix elements
of $\widehat{H}_{DC}$ with respect to a basis of
$n$--electron
configurational states labelled $|\gamma_r J_r\rangle, r =
1,2,3,\ldots$  The present program computes the pure angular
coefficients $t_{rs}^k(ab)$ (only for $k=0$ in the present version)
and $v_{rs}^k(abcd))$ in the
 expression
\begin{equation}
\label{eq:DC-c1}
\D
\langle \gamma_r J_r|\widehat{H}_{DC}|\gamma_s J_s\rangle =
\sum_{ab}\left\{ t^0_{rs}(ab)\,(a|\widehat{H}_D|b) + {\sum_{k}}
\sum_{cd} v^{(k)}_{rs}(abcd)\,X^{(k)}(abcd)\right\},
\end{equation}
where $X^{(k)}(abcd)$ is the effective interaction strength of
the
two--electron interaction with respect to the orbitals concerned. 
Here in (\ref{eq:DC-c1}),
the interaction is the Coulomb potential only,
where
\begin{eqnarray}
\label{eis-c}
X^{(k)}(abcd)\\
& & = (-1)^k \langle  n_a l_a j_a \| C^{(k)} \| n_{c} l_{c}
j_{c} \rangle \langle n_b l_b j_b \| C^{(k)} \|
n_{d} l_{d} j_{d} \rangle R^k(n_a l_a j_a
n_b l_b j_b n_{c} l_{c} j_{c}
n_{d} l_{d} j_{d} ) , \nonumber
\end{eqnarray}
but the formalism can
also be used for the Breit interaction, with a different definition
of
$X^{(k)}(abcd)$ and different selection rules.  
Equation (\ref{eq:DC-c1}) is a rearrangement of the formula used
in
GRASP92~\cite[eqn. (3.10)]{GRASP92},
\begin{eqnarray}
\label{eq:DC-c}
\D
\langle \gamma_r J_r|\widehat{H}_{DC}|\gamma_s J_s\rangle  =
\sum_{ab}\left\{ T_{rs}(ab)I(ab) + {\sum_{k}}
\sum_{cd}
 V^{(k)}_{rs}(abcd)\,R^{(k)}(abcd)\right\},
\end{eqnarray}
where
\[
I(ab) = \left(a|\widehat{H}_D|b \right)
\]
and
\[
R^{(k)}(abcd) =  \left(ab \left|{r_<}^k/{r_>}^{k+1}\right|cd\right)
\]
is a relativistic Slater integral over orbitals $a,b,c,d$
(in the usual notation).  Whilst the one--electron part is the same,
equation (\ref{eq:DC-c}) expands the effective interaction strength as
a traditional sum of Slater integrals, and $V^{(k)}_{rs}(abcd)$
therefore differs from $v^{(k)}_{rs}(abcd)$.  The latter can be
used for \textit{any} two--electron interaction, and it is no
longer
necessary to treat Coulomb and Breit interactions on a different
footing in the manner of GRASP92. Other intractions such as the
lowest--order normal mass shift and the specific mass shift
~\cite[eqns. (3.12--3.14)]{GRASP92} can be handled within the same
scheme. The way in which the coefficients are built up is described
below.
\subsection{One--particle operators}

The matrix elements of a one--particle scalar operator
$\widehat{F}^{(0)}$ between configuration state functions with $u$
open shells can be expressed as a sum over
one--electron contributions
\begin{eqnarray}
\label{eq:one-a}
\D  \left (\psi _u^{bra}( J)\left \| \widehat{F}^{(0)}\right\|
   \psi _u^{ket}( J^{\prime }) \right ) =
    {\sum_{n_i\kappa _i,n_j\kappa _j}}
  \left (\psi _u^{bra}( J) \left\| \widehat{F}
   ( n_i \kappa _i, n_j \kappa _j )
   \right\| \psi _u^{ket}( J^{\prime }) \right)
\end{eqnarray}
where
\begin{eqnarray}
\label{eq:one-b}
\lefteqn{\D
  \left (\psi _u^{bra}( J) \left\| \widehat{F}
   ( n_i \kappa _i, n_j \kappa _j )
   \right\| \psi _u^{ket}( J^{\prime }) \right )}
   \nonumber  \\[1ex]
   & & =
   \D ( -1)^{\Delta +1} \sqrt{2j_i +1}\; R\left( j_i, j_j,\Lambda
  ^{bra},\Lambda ^{ket} \right) \, \delta( \kappa _i , \kappa _j)\,
   \left( n_i\kappa _i\left\|f^{(0) } \right\|n_j\kappa _j \right)
    \nonumber  \\[1ex]
   & & \times  \left\{ \delta ( n_i , n_j )
   \left( j_{i}^{N_i}\;\alpha _i Q_i J_i \left\|
   \left[ a^{\left( q \; \; j_i\right)}_{1/2}
   \times  a^{\left( q \; \; j _i
   \right)} _{-1/2}\right] ^{\left( 0  \right) }
   \right\|j_{i}^{N_{i}} \;\alpha _i Q_i J_i\right)
   \nonumber \right. \\[1ex]
   & & \left. + (1-\delta ( n_i , n_j ))
   \left( j_{i}^{N_i}\,\alpha _i Q_i J_i \left\|
   a^{( q \,j _i)}_{1/2}
   \right\|j_i^{N_{i}^{\prime }} \,\alpha _i Q_i J_i \right)
   \left( j_j^{N_j}\;\alpha _j Q_j J_j \left\|
    a^{( q \, j _j)}_{-1/2}
   \right\|j_j^{N_{j}^{\prime }} \,\alpha_j Q_j J_j \right) 
   \right\}.
\nonumber
\end{eqnarray}
All states are defined in  $jj$--coupling.  $(\psi _u^{bra}\left(
J\right) ||$ and $||\psi _u^{ket}\left( J^{\prime }
\right) )$ are respectively bra and ket functions with $u$ open
subshells, $ \kappa \equiv (2j+1)(l-j)$, $\left(n_i\kappa _i
\left\|f^{( 0 ) }\right\|n_j\kappa _j\right) $ is
the one--electron reduced matrix element of
the operator
$\widehat{F}^{\left( 0 \right)}$,
$\Lambda^{bra} \equiv
\left( J_i,J_j,J_{i^{\prime }},J_{j^{\prime }}\right)^{bra}$
and
$\Lambda^{ket}\equiv
\left( J_i,J_j,J_{i^{\prime }},J_{j^{\prime }}\right)^{ket}$
denote the
respective sets of active subshell angular momenta. The operators
$a^{( q \, j)}_{m_q}$ are second quantization operators
in quasispin space of rank $q = 1/2$. The operator
$a^{( q \, j)}_{1/2 \; m_j} =
a^{\left( j \right) }_{m_j}$
creates
electrons with angular momentum quantum numbers $j,m_j$ and its
conjugate $a^{( q \, j)}_{-1/2 \, m_j} =
\tilde{a}_{m_j }^{( j )}  =
( -1)^{j-m_j }a_{-m_j }^{( j ) +}$ annihilates electrons with the
same quantum numbers  $j,m_j$ in a
given subshell.
\subsubsection{Recoupling matrix}

The recoupling matrix
$R\left( j_i, j_j,\Lambda ^{bra},\Lambda ^{ket} \right)$
in (\ref{eq:one-a}) is particularly simple. It is either a product of
delta functions~\cite[eqn. (18)]{method2} when
$n_i\kappa _i = n_j \kappa _j$
or a combination of delta functions and $6j-$
coefficients~\cite[eqn. (22)]{method2} when
$n_i\kappa _i \neq n_j\kappa _j $.

\subsubsection{Matrix elements of irreducible tensor operators}

By applying the Wigner--Eckart theorem in quasispin
space we obtain the submatrix elements of operators of type
$a_{m_q}^{( q j ) }$ in the
form
\cite{method1}
\begin{eqnarray}
\label{eq:theo-g}
\lefteqn{
   \left( j^N\,\alpha QJ\left\|a_{m_q}^{( q\;j) }\right\|j^{N^{\prime }}
   \,\alpha ^{\prime }Q^{\prime }J^{\prime }\right)
   \nonumber}  \\
  & & \D = -[Q] ^{-1/2}\left[
              \begin{array}{ccc}
                Q^{\prime } & 1/2 & Q \\
               M_Q^{\prime } & m_q & M_Q
              \end{array}
                       \right]
   \left( j^N \,\alpha QJ \ltbar a^{( q \; j)} \rtbar
    j\,\alpha^{\prime} Q^{\prime }J^{\prime}\right),
\end{eqnarray}
where we have used the conventional shorthand
notation
$(2k+1)\cdot
...\equiv [k,...]$, and the last factor is a reduced
coefficient of
fractional parentage. The submatrix elements of the simplest compound
tensor operator of type
$\D\left[ a^{( q \;  j)}_{m_{q2}} \times a^{( q \; j)} _{m_{q2}}
\right]^{( k_j)}$
uses
\begin{eqnarray}
\label{eq:theo-gb}
\lefteqn{
   \left( nj^N\,\alpha QJ \left\|
      \left[ a_{m_{q1}}^{( q \; j) } \times
                 a_{m_{q2}}^{( q \; j) } \right]^{( k_j) }
   \right\|nj^{N^{\prime }}\;\alpha ^{\prime}Q^{\prime}J^{\prime} \right)}
   \nonumber  \\[1ex]
   & & =
   \D {\sum_{ k_q, m_q }}\left[ Q\right]^{-1/2}
     \left[ \begin{array}{ccc}
            q & q & k_q \\
           m_{q1} & m_{q2} & m_q \end{array} \right]
     \left[ \begin{array}{ccc}
           Q^{\prime } & k_q & Q \\
          M_Q^{\prime } & m_q & M_Q \end{array} \right]
   \nonumber  \\[1ex]
   & & \D \times
   \left( nj\;\alpha QJ \ltbar W^{\left( k_q k_j \right)} \rtbar
       nj\;\alpha ^{\prime }Q^{\prime }J^{\prime}\right).
\end{eqnarray}
where $\left( nj\;\alpha QJ \ltbar W^{\left( k_q k_j \right) }
\rtbar nj\;\alpha^{\prime }Q^{\prime }J^{\prime }\right) $ denotes
the
reduced matrix element of the tensor operator $W^{\left( k_q k_j
\right) }\left( nj,nj\right) = \left[
a^{( q\,j) }\times a^{( q\,j) }\right] ^{(k_q k_j) }$ in quasi--spin
space. In terms of the fully reduced coefficients of fractional
parentage $\left(j\,\alpha QJ\ltbar a^{\left(
qj\right)}\rtbar
j\,\alpha^{\prime}Q^{\prime } J^{\prime }\right)$, we
find
\begin{eqnarray}
 \label{eq:theo-gc}
\lefteqn{
   \left( n j\,\alpha Q J \ltbar W^{( k_q k_j)} \rtbar
       n j\,\alpha^{\prime }  Q^{\prime } J^{\prime } \right)}
   \nonumber  \\[1ex]
   & & \D =
   (-1)^{Q+J+Q^{\prime }+J^{\prime}+k_q+k_j}
   \left[k_q,k_j\right]^{1/2}
  \sum_{\alpha^{\prime \prime } Q^{\prime \prime } J^{\prime \prime }}
   \left\{
       \begin{array}{ccc}
            q & q & k_q \\
         Q^{\prime } & Q & Q^{\prime \prime }
            \end{array} \right\}
   \left\{
        \begin{array}{ccc}
           j & j & k_j \\
        J^{\prime } & J & J^{\prime \prime } \end{array} \right\}
   \nonumber  \\[1ex]
   & &\D  \mbox{\hspace{7em}}     \times
   \left( j\,\alpha Q J \ltbar a^{(q\,j)} \rtbar
   j\,\alpha^{\prime\prime} Q^{\prime \prime} J^{\prime \prime} \right)
   \left( j\,\alpha^{\prime \prime }
   Q^{\prime \prime } J^{\prime \prime }\ltbar a^{(q\,j)} \rtbar
    j\,\alpha^{\prime } Q^{\prime} J^{\prime } \right).
    \end{eqnarray}
This construction has the advantage that the completely
reduced matrix
elements on the right hand side of~(\ref{eq:theo-g}) and
(\ref{eq:theo-gb}) are independent of the occupation number of the
shell, and so reduces requirements of storage in comparison with
earlier work.  These formulae are evaluated in
the module
\texttt{rabs\_{}rcfp}~\cite{rabs_rcfp}.
%
%
%

The phase factor arises from the reordering needed to match
the recoupled
creation and annihilation operators in the bra and ket vectors.
We have
\begin{equation}
\label{eq:one-ti}
   \Delta = 0.
\end{equation}
when $n_i\kappa _i = n_j\kappa _j$;
otherwise
\begin{equation}
\label{eq:one-tj}
   \Delta = 1+ \sum_{r=a}^{b-1}
   N_r,
\end{equation}
where $N_r$ is the occupation number of subshell $r$,
$a = \min \{ i,j \} $, and
$b = \max \{ i,j \}$.
\subsubsection{The one--electron submatrix element}

It only remains to define the one--electron interaction matrix
element
\[
\left( n_i\kappa _i \left\|f^{(0)}\right \|n_j\kappa _j\right)
\]
in (\ref{eq:one-a}).  The only operator required in this
implementation is the matrix element of the Dirac operator, a
tensor operator of rank zero, 
\begin{eqnarray}
\label{eq:t-coeff}
   \left( n_i\kappa _i
   \left\|\widehat{H}_D \right\| n_j\kappa _j\right)
   = I(n_il_ij_i,n_il_ij_i)\delta(\kappa_i,\kappa_j),
\end{eqnarray}
where $I(n_il_ij_i,n_il_ij_i)$ is defined
by~\cite[eqn. (22)]{Grant-a}. The Dirac kinetic energy operator,
denoted by $T$ in ~\cite[eqn. (3.13)]{GRASP92}, can be obtained from
this by setting the nuclear charge $Z=0$. The coefficients
$T_{rs}(ab)$ in (\ref{eq:DC-c1}) can now be identified by
inserting (\ref{eq:t-coeff}) in (\ref{eq:one-a}).
Meanwhile the pure angular coefficients $t_{rs}^0(ab)$ for 
one--electron operators can be identified by inserting
\[
\left( n_i\kappa _i \left\|f^{(0)}\right \|n_j\kappa _j\right) = 1
\]
in (\ref{eq:one-a}).

\subsection{Two--particle operators}

According to~\cite{method2}, the matrix element of any two--particle
scalar operator $\widehat{G}^{(kk0)}$ between configuration state
functions with $u$ open shells, can be
written
\begin{eqnarray}
\label{eq:theo-a}
\lefteqn{
   \left(\psi _u^{bra}( J) \left\|\widehat{G}^{(kk0)} \right\|
   \psi _u^{ket}( J^{\prime }) \right)}\nonumber \\[1ex]
   & = & \D {\sum_{n_i \kappa _i,n_j \kappa _j,
   n_{i^{\prime }} \kappa _{i^{\prime }},n_{j^{\prime}}
   \kappa _{j^{\prime }}}}
   \left(\psi _u^{bra}( J) \left\| \widehat{G}
   ( n_i \kappa _i,n_j \kappa _j,n_{i^{\prime }} \kappa _{i^{\prime }},
   n_{j^{\prime }} \kappa _{j^{\prime }} )
   \right\|\psi _u^{ket}( J^{\prime })
   \right)
\end{eqnarray}
where
\begin{eqnarray}
\label{eq:theo-a1}
\lefteqn{
   (\psi _u^{bra}\left( J\right) || \widehat{G}
   \left( n_i \kappa _i,n_j \kappa _j,n_{i^{\prime }}
   \kappa _{i^{\prime }},
   n_{j^{\prime }} \kappa _{j^{\prime }} \right) ||
   \psi _u^{ket}\left( J^{\prime } \right) )  \nonumber } \\[1ex]
   & & =
   \D    {\sum_{k_{12}}} (-1)^\Delta \Theta^{\prime }
   ( n_i l_i j_i,n_j l_j j_j,n_{i^{\prime }} l_{i^{\prime }}
   j_{i^{\prime}},n_{j^{\prime }} l_{j^{\prime }} j_{j^{\prime }},\Xi)
   \nonumber \\[1ex]
   & & \times  T\left(
   n_i j_i,n_j j_j,n_{i^{\prime }} j_{i^{\prime }},n_{j^{\prime }}
   j_{j^{\prime }}, \Lambda^{bra}, \Lambda^{ket}, \Xi ,\Gamma \right)
   R\left( j_i, j_j, j_{i^{\prime }}, j_{j^{\prime }},
   \Lambda^{bra}, \Lambda ^{ket}, \Gamma \right),
 \nonumber
\end{eqnarray}
$\Gamma$ specifies the recoupling scheme required for each matrix
element and
$\Xi$, when required, specifies the coupling scheme of the tensor
operators defining each matrix element. The operator
$\widehat{G}^{(kk0)}$ couples tensor operators of rank $k$ for each
electron to give an overall scalar operator.
\begin{table}
\begin{center}
\caption{Scheme of the definitions for matrix elements of any
two--particle operator. The operators a, W, aW and Wa
defined in
(\ref{eq:theo-b}) -- (\ref{eq:theo-f}) act on the
indicated subshells.}
\vspace{3 mm}
\label{israis}
\begin{small}

\begin{tabular}{|r|cccc|l|l|cccc|l|} \hline
{\bf No.}
& $i$ & $j$ & $i^{\prime }$ & $j^{\prime }$
& $\Gamma$  & $\Xi$
& $\alpha$  & $\beta$
& $\gamma$  & $\delta$ & $\varphi$ \\ \hline
{\bf 1.}     & $\alpha$  & $\alpha$ & $\alpha$ & $\alpha$
& --                                           & $j_{\alpha}, k$
& WW & --              & --              & --
& --                                                           \\
             &           &          &          &
& --                                           & $j_{\alpha}$
& W & --              & --              & --
& --                                                           \\
{\bf 2.}     & $\alpha$  & $\beta$  & $\alpha$ & $\beta$
& $k$                                          & $j_{\alpha}, j_{\beta}$
& W & W & --              & --
& $0$                                                          \\
{\bf 3.}     & $\beta$   & $\alpha$ & $\beta$  & $\alpha$
& $k$                                          & $j_{\alpha}, j_{\beta}$
& W & W & --              & --
& $0$                                                          \\
{\bf 4.}     & $\alpha$  & $\beta$  & $\beta$  & $\alpha$
& $k_{12}$                                     & $j_{\alpha}, j_{\beta}$
& W & W & --              & --
& $0$                                                           \\
{\bf 5.}     & $\beta$   & $\alpha$ & $\alpha$ & $\beta$
& $k_{12}$                                     & $j_{\alpha}, j_{\beta}$
& W & W & --              & --
& $0$                                                           \\
{\bf 6.}     & $\alpha$  & $\alpha$ & $\beta$  & $\beta$
& $k_{12}$                                     & $j_{\alpha}, j_{\beta}$
& W & W & --              & --
& $j_{\alpha}+j_{\beta}+k_{12}$                                 \\
{\bf 7.}     & $\beta$   & $\alpha$ & $\alpha$ & $\alpha$
& $j_{\beta}$                                  & $j_{\alpha}, k_{12}$
& aW & a & --              & --
& $k+k_{12}$                                                    \\
{\bf 8.}     & $\alpha$  & $\beta$  & $\alpha$ & $\alpha$
& $j_{\beta}$                                  & $j_{\alpha}, k_{12}$
& aW & a & --              & --
& $k$                                                           \\
{\bf 9.}     & $\beta$   & $\beta$  & $\beta$  & $\alpha$
& $j_{\alpha}$                                 & $j_{\beta}, k_{12}$
& a & Wa & --              & --
& $k+k_{12}$                                                    \\
{\bf 10.}    & $\beta$   & $\beta$  & $\alpha$ & $\beta$
& $j_{\alpha}$                                 & $j_{\beta}, k_{12}$
& a & Wa & --              & --
& $k$                                                           \\
{\bf 11.}    & $\beta$   & $\gamma$ & $\alpha$ & $\gamma$
& $j_{\alpha}, j_{\beta}, k$                   & --
& a & a & W & --
& $1+j_{\alpha}+j_{\beta}-k$                                    \\
{\bf 12.}    & $\gamma$  & $\beta$  & $\gamma$ & $\alpha$
& $j_{\alpha}, j_{\beta}, k$                   & --
& a & a & W & --
& $1+j_{\alpha}+j_{\beta}-k$                                    \\
{\bf 13.}    & $\gamma$  & $\beta$  & $\alpha$ & $\gamma$
& $j_{\alpha}, j_{\beta}, k_{12}$              & --
& a & a & W & --
& $1+j_{\alpha}+j_{\beta}-k_{12}$                               \\
{\bf 14.}    & $\beta$   & $\gamma$ & $\gamma$ & $\alpha$
& $j_{\alpha}, j_{\beta}, k_{12}$              & --
& a & a & W & --
& $1+j_{\alpha}+j_{\beta}-k_{12}$                               \\
{\bf 15.}    & $\gamma$  & $\gamma$ & $\alpha$ & $\beta$
& $j_{\alpha}, j_{\beta}, k_{12}$              & --
& a & a & W & --
& $j_{\alpha}+j_{\gamma}+k_{12}$                                \\
{\bf 16.}    & $\gamma$  & $\gamma$ & $\beta$  & $\alpha$
& $j_{\alpha}, j_{\beta}, k_{12}$              & --
& a & a & W & --
& $j_{\alpha}+j_{\gamma}$                                       \\
{\bf 17.}    & $\alpha$  & $\beta$  & $\gamma$ & $\gamma$
& $j_{\alpha}, j_{\beta}, k_{12}$              & --
& a & a & W & --
& $j_{\alpha}+j_{\beta}+k_{12}$                                 \\
{\bf 18.}    & $\beta$   & $\alpha$ & $\gamma$ & $\gamma$
& $j_{\alpha}, j_{\beta}, k_{12}$              & --
& a & a & W & --
& $j_{\beta}+j_{\gamma}$                                        \\
{\bf 19.}    & $\alpha$  & $\beta$  & $\gamma$ & $\delta$
& $j_{\alpha}, j_{\beta}, j_{\gamma}, j_{\delta}$ & --
& a & a & a & a
& $j_{\alpha}+j_{\beta}+k_{12}$                                 \\
{\bf 20.}    & $\beta$   & $\alpha$ & $\gamma$ & $\delta$
& $j_{\alpha}, j_{\beta}, j_{\gamma}, j_{\delta}$ & --
& a & a & a & a
& $j_{\beta}+j_{\gamma}+k_{12}$                                 \\
{\bf 21.}    & $\alpha$  & $\beta$  & $\delta$ & $\gamma$
& $j_{\alpha}, j_{\beta}, j_{\gamma}, j_{\delta}$ & --
& a & a & a & a
& $j_{\beta}+j_{\gamma}$                                        \\
{\bf 22.}    & $\beta$   & $\alpha$ & $\delta$ & $\gamma$
& $j_{\alpha}, j_{\beta}, j_{\gamma}, j_{\delta}$ & --
& a & a & a & a
& $j_{\beta}+j_{\gamma}$                                        \\
{\bf 23.}    & $\gamma$  & $\delta$ & $\alpha$ & $\beta$
& $j_{\alpha}, j_{\beta}, j_{\gamma}, j_{\delta}$ & --
& a & a & a & a
& $j_{\alpha}+j_{\delta}+k_{12}$                                \\
{\bf 24.}    & $\gamma$  & $\delta$ & $\beta$  & $\alpha$
& $j_{\alpha}, j_{\beta}, j_{\gamma}, j_{\delta}$ & --
& a & a & a & a
& $j_{\alpha}+j_{\delta}+k_{12}$                                \\
{\bf 25.}    & $\delta$  & $\gamma$ & $\alpha$ & $\beta$
& $j_{\alpha}, j_{\beta}, j_{\gamma}, j_{\delta}$ & --
& a & a & a & a
& $j_{\alpha}+j_{\delta}$                                       \\
{\bf 26.}    & $\delta$  & $\gamma$ & $\beta$  & $\alpha$
& $j_{\alpha}, j_{\beta}, j_{\gamma}, j_{\delta}$ & --
& a & a & a & a
& $j_{\alpha}+j_{\delta}$                                       \\
{\bf 27.}    & $\alpha$  & $\gamma$ & $\beta$  & $\delta$
& $j_{\alpha}, j_{\beta}, j_{\gamma}, j_{\delta}$ & --
& a & a & a & a
& $0$                                                           \\
{\bf 28.}    & $\alpha$  & $\gamma$ & $\delta$ & $\beta$
& $j_{\alpha}, j_{\beta}, j_{\gamma}, j_{\delta}$ & --
& a & a & a & a
& $0$                                                           \\
{\bf 29.}    & $\gamma$  & $\alpha$ & $\delta$ & $\beta$
& $j_{\alpha}, j_{\beta}, j_{\gamma}, j_{\delta}$ & --
& a & a & a & a
& $0$                                                           \\
{\bf 30.}    & $\gamma$  & $\alpha$ & $\beta$  & $\delta$
& $j_{\alpha}, j_{\beta}, j_{\gamma}, j_{\delta}$ & --
& a & a & a & a
& $0$                                                           \\
{\bf 31.}    & $\beta$   & $\delta$ & $\alpha$ & $\gamma$
& $j_{\alpha}, j_{\beta}, j_{\gamma}, j_{\delta}$ & --
& a & a & a & a
& $j_{\alpha}+j_{\beta}+j_{\gamma}+j_{\delta}$                  \\
{\bf 32.}    & $\delta$  & $\beta$  & $\gamma$ & $\alpha$
& $j_{\alpha}, j_{\beta}, j_{\gamma}, j_{\delta}$ & --
& a & a & a & a
& $j_{\alpha}+j_{\beta}+j_{\gamma}+j_{\delta}$                  \\
{\bf 33.}    & $\beta$   & $\delta$ & $\gamma$ & $\alpha$
& $j_{\alpha}, j_{\beta}, j_{\gamma}, j_{\delta}$ & --
& a & a & a & a
& $j_{\alpha}+j_{\beta}+j_{\gamma}+j_{\delta}$                  \\
{\bf 34.}    & $\delta$  & $\beta$  & $\alpha$ & $\gamma$
& $j_{\alpha}, j_{\beta}, j_{\gamma}, j_{\delta}$ & --
& a & a & a & a
& $j_{\alpha}+j_{\beta}+j_{\gamma}+j_{\delta}$                 \\
{\bf 35.}    & $\alpha$  & $\delta$ & $\beta$  & $\gamma$
& $j_{\alpha}, j_{\beta}, j_{\gamma}, j_{\delta}$ & --
& a & a & a & a
& $1+j_{\gamma}+j_{\delta}-k$                                  \\
{\bf 36.}    & $\delta$  & $\alpha$ & $\gamma$ & $\beta$
& $j_{\alpha}, j_{\beta}, j_{\gamma}, j_{\delta}$ & --
& a & a & a & a
& $1+j_{\gamma}+j_{\delta}-k$                                  \\
{\bf 37.}    & $\alpha$  & $\delta$ & $\gamma$ & $\beta$
& $j_{\alpha}, j_{\beta}, j_{\gamma}, j_{\delta}$ & --
& a & a & a & a
& $1+j_{\gamma}+j_{\delta}-k_{12}$                             \\
{\bf 38.}    & $\delta$  & $\alpha$ & $\beta$  & $\gamma$
& $j_{\alpha}, j_{\beta}, j_{\gamma}, j_{\delta}$ & --
& a & a & a & a
& $1+j_{\gamma}+j_{\delta}-k_{12}$                             \\
{\bf 39.}    & $\beta$   & $\gamma$ & $\alpha$ & $\delta$
& $j_{\alpha}, j_{\beta}, j_{\gamma}, j_{\delta}$ & --
& a & a & a & a
& $1+j_{\alpha}+j_{\beta}-k$                                   \\
{\bf 40.}    & $\gamma$  & $\beta$  & $\delta$ & $\alpha$
& $j_{\alpha}, j_{\beta}, j_{\gamma}, j_{\delta}$ & --
& a & a & a & a
& $1+j_{\alpha}+j_{\beta}-k$                                   \\
{\bf 41.}    & $\beta$   & $\gamma$ & $\delta$ & $\alpha$
& $j_{\alpha}, j_{\beta}, j_{\gamma}, j_{\delta}$ & --
& a & a & a & a
& $1+j_{\alpha}+j_{\beta}-k_{12}$                             \\
{\bf 42.}    & $\gamma$  & $\beta$  & $\alpha$ & $\delta$
& $j_{\alpha}, j_{\beta}, j_{\gamma}, j_{\delta}$ & --
& a & a & a & a
& $1+j_{\alpha}+j_{\beta}-k_{12}$       \\
\hline
\end{tabular}
\end{small}
\end{center}
\end{table}

\medskip

From (\ref{eq:theo-a}) we see that the matrix element of
any
two--particle
operator can be written as a sum over all possible sets of active
shell quantum numbers $n_i \kappa _i$, $n_j \kappa _j$,
$n_{i^{\prime }} \kappa _{i^{\prime }}$, $n_{j^{\prime }} \kappa
_{j^{\prime }}$. The systematic analysis of \cite{method2} aims to
minimize the
computation needed in this expansion.  The parameter
distributions are
presented in Table~\ref{israis}. Note that for distributions 2--5 and
19--42 the subshell labels are ordered so that $\alpha < \beta <
\gamma < \delta$, while for distributions 6--18 no conditions
upon the
ordering are imposed.  We discuss these structures in more
detail below.
\subsubsection{Recoupling matrix}

The recoupling coefficients defined in~\cite{MCP,MCP75,MCBP} did not
reduce the recoupling coefficients to their simplest forms but relied
on the analysis module of the NJSYM package (later NJGRAF) to perform
the reduction mechanically.  The analysis~\cite{method2} leads to
simpler forms denoted by $R\left( j_i, j_j, j_{i^{\prime }},
j_{j^{\prime }},\Lambda^{bra},\Lambda ^{ket},\Gamma \right)$.
In the case of one interacting shell
$R\left( j_i, j_j, j_{i^{\prime }}, j_{j^{\prime }},\Lambda
^{bra},\Lambda ^{ket},\Gamma \right) $
reduces to delta functions~\cite[eqn. (18)]{method2}.
For two, three and four interacting shells, the recoupling
coefficients are given by~\cite[eqns. (22), (23)
and (24)]{method2},
replacing $l,L$ by $j,J$ respectively. The recoupling parameters
$\Gamma$ for each distribution can be found in
Table~\ref{israis}.
\subsubsection{Matrix elements of irreducible tensor operators}

The expressions
$T\left( n_i j_i,n_j j_j,n_{i^{\prime }} j_{i^{\prime}},
n_{j^{\prime }} j_{j^{\prime }}, \Lambda ^{bra}, \Lambda ^{ket},\Xi,
\Gamma \right)$ are matrix elements of
standard subshell
creation/annihilation operators
\begin{eqnarray}
\label{eq:theo-b}
   a & = & a_{m_q}^{\left( q j \right) }, \\
\label{eq:theo-c}
   W & = & \left[ a_{m_{q1}}^{\left( q j \right) }\times
   a_{m_{q2}}^{\left( q j \right) }\right] ^{\left( k_{12} \right) },
   \\
\label{eq:theo-d}
   aW & = & \left[ a_{m_{q1}}^{\left( q j \right) }\times
   \left[
   a_{m_{q2}}^{\left( q j \right) }\times a_{m_{q3}}^{\left( q j \right) }
   \right] ^{\left( k_{12} \right) }\right] ^{\left( k_2 \right) },
   \\
\label{eq:theo-e}
   Wa & = & \left[ \left[ a_{m_{q1}}^{\left( q j \right)}
   \times
   a_{m_{q2}}^{\left( q j \right) }\right] ^{\left( k_{12} \right) }
   \times a_{m_{q3}}^{\left( q j \right) }\right] ^{\left( k_2\right) },
   \\
\label{eq:theo-f}
   WW & = & \left[ \left[ a_{m_{q1}}^{\left( q j \right)}
   \times
   a_{m_{q2}}^{\left( q j \right) }\right] ^{\left( k
   \right) }\times \left[ a_{m_{q3}}^{\left( q j \right) }\times
   a_{m_{q4}}^{\left( q j \right) }\right] ^{\left( k
   \right) }\right] ^{\left( 0 \right)} .
\end{eqnarray}

The creation and annihilation operators
in (\ref{eq:theo-b})--(\ref{eq:theo-f}) refer to a
single subshell.
The evaluation of the submatrix elements of operators
of type
$a$~(\ref{eq:theo-b}) and the simplest compound tensor operator of
type $W$~(\ref{eq:theo-c}) was explained in section~2.1.2.
For types (\ref{eq:theo-d})--(\ref{eq:theo-f}), we use
the formula
\begin{eqnarray}
\label{eq:theo-h}
\lefteqn{
   \left(nj^N\,\alpha QJ \left\|
      \left[ U^{( k_1)}(nj) \times V^{( k_2 )}(nj) \right]^{(k)}
        \right\| nj^{N^{\prime }}\,\alpha^{\prime }
                      Q^{\prime } J^{\prime }\right)}
    \\[1ex]
& & = (-1)^{J+J^{\prime }+k}[k]^{1/2}
   \D {\sum_{\alpha ^{\prime \prime } Q^{\prime \prime }
             J^{\prime\prime }}}
      \left\{
        \begin{array}{ccc}
           k_1 & k_2 & k \\
        J^{\prime } & J & J^{\prime \prime }
           \end{array} \right\},
   \nonumber  \\
&  & \times\,
    \left(nj^N\,\alpha Q J \left\| U^{(k_1)}(nj) \right\|
    nj^{N^{\prime \prime }}\, \alpha^{\prime \prime } Q^{\prime\prime}
                 J^{\prime \prime }\right)
    \left(nj^{N^{\prime \prime }}\,\alpha ^{\prime \prime }
   Q^{\prime \prime } J^{\prime \prime }
        \left\| V^{( k_2 )}(nj) \right\|
     nj^{N^{\prime }}\,\alpha^{\prime } Q^{\prime} J^{\prime } \right)
   \nonumber
\end{eqnarray}
where $U^{\left( k_1\right) }\left( n j \right),
V^{( k_2 )}\left( n j \right)$ are either of type (\ref{eq:theo-b})
or type (\ref{eq:theo-c}).
The occupation number $N^{\prime \prime }$ is defined by
second quantization operators occurring in $U^{\left( k_1
\right) }\left( n j \right) $ and $V^{( k_2)}\left(n j \right) $.
The module \texttt{rabs\_{}rcfp}~\cite{rabs_rcfp} performs the
evaluation of these formulae.
\subsubsection{Phase factors $\Delta $}
These arise from the reordering necessary to match the recoupled
creation and annihilation operators in bra and ket vectors
contributing to the matrix element.  For each of the cases considered
in Table~\ref{israis} we find\\[1ex]
\textit{Cases 1--6}:
\begin{equation}
\label{eq:ti}
   \Delta = 0.
\end{equation}
\textit{Cases 7--18}:
\begin{equation}
\label{eq:tj}
   \Delta = 1+\sum_{r=
   i}^{j-1}N_r,
\end{equation}
where $N_r$ is the occupation number of subshell $r$. If $\alpha
<\beta $, then $i= \alpha ,$
$j= \beta $, and if $\alpha >\beta $,
then
$i= \beta ,$ $j= \alpha$.\\[1ex]
\textit{Cases 19--42}:
\begin{equation}
\label{eq:tl}
   \Delta = \sum_{k=
   \alpha }^{\beta -1}N_k+
   \sum_{k= \gamma}^{\delta
   -1}N_k.
\end{equation}
\subsubsection{The coefficients $\Theta ^{\prime }\left( n_i l_i j_i,
n_j l_j j_j,n_{i^{\prime }} l_{i^{\prime }} j_{i^{\prime }},
n_{j^{\prime}} l_{j^{\prime }} j_{j^{\prime }}, \Xi \right) $}

The effective interaction strength of order $k$ of a two--electron
operator, is denoted by
\[
\left( n_i l_i j_i n_j l_j
j_j\left\| g^{( k k )} \right\| n_{i^{\prime }} l_{i^{\prime }}
j_{i^{\prime }} n_{j^{\prime}} l_{j^{\prime }} j_{j^{\prime }}
\right)
\] in ~\cite{method6}  and by
\[
X^k\left( n_i l_i j_i, n_j l_j j_j,
n_{i^{\prime }} l_{i^{\prime }} j_{i^{\prime }}, n_{j^{\prime}}
l_{j^{\prime }} j_{j^{\prime }} \right)
\]
 in ~\cite{Grant-a}.
The coefficients $\Theta ^{\prime }\left( n_i l_i j_i,
n_j l_j j_j, n_{i^{\prime }} l_{i^{\prime }} j_{i^{\prime }},
n_{j^{\prime}} l_{j^{\prime }} j_{j^{\prime }}, \Xi \right) $
for the different cases tabulated in Table \ref{israis}) have
different multiplicative factors defined as follows:

\medskip

\textit{Case 1}: Single subshell ($\alpha \alpha \alpha
 \alpha$)
\begin{eqnarray}
\label{eq:theta-a}
\lefteqn{
   \Theta ^{\prime}_{IIa}\left( n_\alpha l_\alpha j_\alpha ,n_\alpha
   l_\alpha
   j_\alpha , n_\alpha l_\alpha j_\alpha , n_\alpha l_\alpha j_\alpha ,\Xi
   \right)}
   \nonumber  \\[1ex]
   & & = \frac{1}{2}[k]^{-1/2}
   \left( n_\alpha l_\alpha j_\alpha n_\alpha l_\alpha j_\alpha \left\|
   g^{\left(k k\right) }
   \right\| n_\alpha l_\alpha j_\alpha n_\alpha l_\alpha j_\alpha \right)
   \delta \left( k_{12},k \right)
\end{eqnarray}
and
\begin{eqnarray}
\label{eq:theta-b}
\lefteqn{
   \Theta ^{\prime} _{IIb}\left( n_\alpha l_\alpha j_\alpha ,n_\alpha
   l_\alpha
   j_\alpha , n_\alpha l_\alpha j_\alpha , n_\alpha l_\alpha j_\alpha ,\Xi
   \right)}
   \nonumber  \\[1ex]
   & & = \left( -1\right) ^{k} \left[ j_{\alpha } \right]
   ^{-1/2}
   \left( n_\alpha l_\alpha j_\alpha n_\alpha l_\alpha j_\alpha \left\|
   g^{\left(k k\right) }
   \right\| n_\alpha l_\alpha j_\alpha n_\alpha l_\alpha j_\alpha \right)
   \delta \left( k_{12},0 \right).
\end{eqnarray}

\textit{Cases 2, 3, 11, 12, 27, 29, 31, 32, 35, 36, 39, 40}:
Subshell assignments $\alpha\beta\alpha\beta,
\beta\alpha\beta\alpha$,
$\beta \gamma \alpha \gamma $, $\gamma \beta \gamma \alpha $,
$\alpha \gamma \beta \delta $, $\gamma \alpha \delta \beta $,
$\beta \delta \alpha \gamma $, $\delta \beta \gamma \alpha $,
$\alpha \delta \beta \gamma $, $\delta \alpha \gamma \beta $,
$\beta \gamma \alpha \delta $, $\gamma \beta \delta \alpha $
\begin{eqnarray}
\label{eq:theta-c}
\lefteqn{
   \Theta ^{\prime }\left(n_i l_i j_i,n_j l_j j_j,
   n_{i^{\prime }} l_{i^{\prime }} j_{i^{\prime }},
   n_{j^{\prime}} l_{j^{\prime }} j_{j^{\prime }}, \Xi \right)}
   \nonumber  \\[1ex]
   & = & \left(-1\right) ^{\varphi} \frac 12\left[
   k\right]
   ^{-1/2} \left( n_i l_i j_i n_j l_j j_j\left\| g^{( k k )} \right\|
   n_{i^{\prime }} l_{i^{\prime }} j_{i^{\prime }} n_{j^{\prime}}
   l_{j^{\prime }} j_{j^{\prime }} \right) \delta \left( k_{12},k
   \right).
\end{eqnarray}
\textit{Cases 6, 15--26}:  Subshell assignments $\alpha \alpha \beta
\beta $, $\gamma \gamma \alpha \beta $,
$\gamma \gamma \beta  \alpha $, $\alpha \beta  \gamma  \gamma $,
$\beta  \alpha \gamma \gamma $, $\alpha \beta  \gamma  \delta $,
$\beta  \alpha \delta \gamma $, $\alpha \beta  \delta  \gamma $,
$\beta  \alpha \gamma \delta $, $\gamma \delta \alpha  \beta  $,
$\delta \gamma \beta  \alpha $, $\gamma \delta \beta   \alpha $,
$\delta \gamma \alpha \beta $
\begin{eqnarray}
\label{eq:theta-d}
\lefteqn{
   \Theta ^{\prime }\left(n_i l_i j_i,n_j l_j j_j,
   n_{i^{\prime }} l_{i^{\prime }} j_{i^{\prime }},
   n_{j^{\prime}} l_{j^{\prime }} j_{j^{\prime }}, \Xi \right)}
   \nonumber  \\[1ex]
   & & = \left(-1\right) ^{1+k+\varphi} \frac 12\left[
   k_{12} \right]
   ^{1/2}
   \left\{
   \begin{array}{ccc}
   j_i & j_{i^{\prime }} & k \\
   j_{j^{\prime }} & j_j & k_{12}
   \end{array}
   \right\}
   \left( n_i l_i j_i n_j l_j j_j\left\| g^{( k k )} \right\|
   n_{i^{\prime }} l_{i^{\prime }} j_{i^{\prime }} n_{j^{\prime}}
   l_{j^{\prime }} j_{j^{\prime }} \right).
\end{eqnarray}
\textit{Cases 4, 5, 7, 8, 9, 10, 13, 14, 28, 30, 33, 34, 37, 38, 41,
42}: Subshell arrangements $\alpha \beta \beta \alpha $, $\beta
\alpha\alpha \beta $, $\beta \alpha \alpha \alpha $,
$\alpha \beta \alpha \alpha $,
$\beta \beta  \beta  \alpha $, $\beta  \beta \alpha \beta  $,
$\gamma \beta \alpha \gamma $, $\beta \gamma \gamma \alpha $,
$\alpha \gamma \delta \beta $, $\gamma \alpha \beta \delta $,
$\beta \delta \gamma \alpha $, $\delta \beta \alpha \gamma $,
$\alpha \delta \gamma \beta $, $\delta \alpha \beta \gamma $,
$\beta \gamma \delta \alpha $, $\gamma \beta \alpha \delta $
\begin{eqnarray}
\label{eq:theta-e}
\lefteqn{
   \Theta ^{\prime }\left(n_i l_i j_i,n_j l_j j_j,
   n_{i^{\prime }} l_{i^{\prime }}j_{i^{\prime }},
   n_{j^{\prime}} l_{j^{\prime }} j_{j^{\prime }}, \Xi \right)}
   \nonumber  \\[1ex]
   & & = \left(-1\right) ^{\varphi} \frac 12\left[ k_{12}
   \right]
   ^{1/2}
   \left\{
   \begin{array}{ccc}
   j_i & j_{i^{\prime }} & k \\
   j_j & j_{j^{\prime }} & k_{12}
   \end{array}
   \right\}
   \left( n_i l_i j_i n_j l_j j_j\left\| g^{( k k )} \right\|
   n_{i^{\prime }} l_{i^{\prime }} j_{i^{\prime }} n_{j^{\prime}}
   l_{j^{\prime }} j_{j^{\prime }} \right).
\end{eqnarray}
The phase factors $\varphi$ in expressions (\ref{eq:theta-c})
 --
(\ref{eq:theta-e}) are defined in column $\varphi$
of Table~\ref{israis}.
This construction exploits the common tensorial structure of
any scalar two--electron operators as 
the Coulomb, Breit and Gaunt interactions
~\cite{Grant-a}) and exploits this similarity to simplify the
calculation of spin--angular coefficients.
The relativistic $jj$--coupling expressions for the
effective interaction
strength of the Coulomb interaction ~\cite[eqn. (86)]{Grant-a}
is
\begin{eqnarray}
\label{eis-c}
\lefteqn{\left( n_i l_i j_i n_j l_j
j_j\left\| g^{( k k )} \right\| n_{i^{\prime }} l_{i^{\prime }}
j_{i^{\prime }} n_{j^{\prime}} l_{j^{\prime }} j_{j^{\prime }}
\right)} \\
& & = (-1)^k \langle  n_i l_i j_i \| C^{(k)} \|
n_{i^{\prime }}
l_{i^{\prime }}
j_{i^{\prime }} \rangle \langle n_j l_j j_j \| C^{(k)} \|
n_{j^{\prime}} l_{j^{\prime }} j_{j^{\prime }} \rangle R^k(n_i l_i j_i
n_j l_j j_j n_{i^{\prime }} l_{i^{\prime }} j_{i^{\prime }}
n_{j^{\prime}} l_{j^{\prime }} j_{j^{\prime }} ) .
\nonumber
\end{eqnarray}

We can now identify the coefficients $V_{rs}^{k}(abcd)$ of (\ref{eq:DC-c1})
by substituting the results above in (\ref{eq:theo-a}). The same
construction can be used for the Gaunt interaction (the leading part
of the magnetic Breit interaction) ~\cite[eqn. (91)]{Grant-a}  and for
the full transverse Breit interaction  ~\cite[eqn. (101)]{Grant-a},
although the selection rules and the effective interaction strengths
corresponding to (\ref{eq:theta-e}) are, of course, different.
The pure angular coefficients $v_{rs}^{k}(abcd)$ for two--electron operators
are the same for all these operators
since the $v_{rs}^{k}(abcd)$  can be identified 
by inserting
\[
\left( n_i l_i j_i n_j l_j
j_j\left\| g^{( k k )} \right\| n_{i^{\prime }} l_{i^{\prime }}
j_{i^{\prime }} n_{j^{\prime}} l_{j^{\prime }} j_{j^{\prime }}
\right) = 1
\]
in (\ref{eq:theo-a}).

\section{Program organization}

\subsection{Overview of the program}

The program ANCO constructs the pure angular coefficients
$t^0_{rs}(ab)$ for one--electron operators and the 
$v^k_{rs}(abcd)$
coefficients contributing to matrix elements of the
Dirac--Coulomb--Breit
Hamiltonian.  The coefficients $T_{rs}(ab)$ and
$V^k_{rs}(abcd)$ used in GRASP92 and earlier version of the system
are available as an option.  The new format generates what we
have called ``pure'' angular momentum coefficients which can be
used
unchanged with any one--particle tensor operator of rank 0, and any
two--particle
interaction. The Coulomb and Breit interactions use different 
subsets of the
complete set of $v^k_{rs}(abcd)$ coefficients, which 
are selected
automatically when multiplying by the relevant effective 
interaction
strengths to complete the matrix element calculation.  The MCP and
MCBP modules of GRASP92 calculated the full matrix elements for each
of these subsets, so that the new formulation reduces the
computational overheads and the memory requirements, which renders
ANCO more suitable for large scale problems. 

\medskip

There are two new modules
\texttt{rabs\_{}recoupling} and \texttt{rabs\_{}anco} for 
extracting
spin--angular coefficients relating to formula (\ref{eq:theo-a}).
The module \texttt{rabs\_{}recoupling} evaluates recoupling
coefficients $R\left( j_i, j_j,\Lambda ^{bra},\Lambda ^{ket} \right)$ 
and
$R\left( j_i, j_j, j_{i^{\prime }}, j_{j^{\prime }},\Lambda
^{bra},\Lambda ^{ket},\Gamma \right)$ as described in~\cite{method2},
module \texttt{rabs\_{}rcfp}~\cite{rabs_rcfp} evaluates the
$T\left( n_i j_i,n_j j_j,n_{i^{\prime }} j_{i^{\prime
}},n_{j^{\prime }}  j_{j^{\prime }},\Lambda^{bra},\Lambda^{ket},\Xi
,\Gamma \right)$ whilst \texttt{rabs\_{}anco} evaluates
all the contributions to (\ref{eq:theo-a}) for both scalar one--
and
two--particle operators.

\medskip

The program ANCO can be run in two modes. The interactive mode is
intended for testing the program and for performing short
calculations involving a small number of configurations.  Normally
the modules  \texttt{rabs\_{}anco} and \texttt{rabs\_{}recoupling}
will be interfaced to GRASP92 or to some other program with 
compatible data
structure to perform multiconfiguration or
configuration--interaction relativistic calculations. 
The program can
handle any combination of open subshells with $j \leq 9/2$, but
subshells with $j > 9/2$ are only allowed if they contain not more
than two electrons. 

\medskip

ANCO is written in Fortran 90/95 and is designed as  an addition to
the RATIP package  \cite{Fritzsche/CFF/Dong:99}. The new Fortran
90/95 standard enables us to define new derived data
types which will enable us to incorporate this module in our
continuing development of large--scale computations for open--shell
atoms and ions.  The full definition of the various derived structures
can be found in the module header of \texttt{rabs\_{}anco}. Here we
need only those types which concern the program output.

\medskip

All information about pure spin--angular coefficients of
scalar one-- and two--particle operators ($t_{rs}(ab)$ and 
$v_{rs}^{k}(abcd)$
coefficients) of the Hamiltonian matrix or of some part of it is 
summarized in the derived data
\texttt{anco\_{}pair\_{}list}.
\begin{verbatim}
   type(anco_csf_pair), dimension(:), pointer :: anco_pair_list
\end{verbatim}

\medskip

which is defined by
\begin{verbatim}
   type :: anco_csf_pair
      integer          :: r, s
      integer          :: no_t_coeff, no_v_coeff
      type(anco_t_coeff), dimension(:), pointer :: t_coeff
      type(anco_v_coeff), dimension(:), pointer :: v_coeff
   end type anco_csf_pair
\end{verbatim}
The integers \texttt{r} and \texttt{s} respectively index the bra--
and the ket-- configuration state functions (CSF) for the current
matrix element.  The variable \texttt{no\_{}t\_{}coeff} is the number
of pure spin--angular coefficients of one--particle operators that can
be constructed for the pair \texttt{r,s}, and the variable
\texttt{no\_{}v\_{}coeff}  gives corresponding data for two--particle
matrix elements. The array \texttt{anco\_{}v\_{}coeff} contains 
pure
spin--angular coefficients for two--particle scalar operators. It is
defined by
\begin{verbatim}
   type :: anco_v_coeff
      integer          :: k
      type(nkappa)     :: a, b, c, d
      real(kind=dp)    :: v
   end type anco_v_coeff
\end{verbatim}
where \texttt{k} is the rank $k$ of the effective
interaction strength,  \texttt{a, b, c, d} point to the relevant
subshells $n_i l_i j_i$, $n_j l_j j_j$, $n_{i^{\prime }} l_{i^{\prime
}} j_{i^{\prime }}$,  $n_{j^{\prime}} l_{j^{\prime }} j_{j^{\prime
}}$, and the pure spin--angular coefficient itself is given
in \texttt{v}. The array \texttt{anco\_{}t\_{}coeff} is defined in
the same way.

\medskip

Memory for the array \texttt{anco\_{}pair\_{}list} is allocated
dynamically using the 
\begin{verbatim}
   allocate( anco_pair_list(1:number_of_pair_list_max))
\end{verbatim}
instruction, and can be deallocated subsequently.
\subsection{Interactive calculations}

A typical interactive dialog for calculating spin--angular
coefficients is shown in Fig.~1.

\medskip

\texttt{Enter a file name for the  anco.sum  file:} 

\medskip
After this prompt, the user should insert the output file name to which the
main output data will be written. This must be followed by the name of
the input file listing CSF in the GRASP92 format. The next
question

\medskip

\texttt{Generate only not trivial angular coefficients
which include (at least one)\\ open shells ?}

\medskip

The response \texttt{y} or \texttt{Y} will cause the program to
calculate coefficients for peel shells only; the reponse
\texttt{n} or
\texttt{N} the program calculate will yield data for all shells (open
and closed).

\medskip

The question

\medskip

\texttt{Generate one-electron angular coefficients for scalar
interactions ?}

\medskip

This requires answer \texttt{n} or \texttt{N} if one--electron
coefficients are not needed.  If the user responds \texttt{y} or
\texttt{Y}, then the prompt

\medskip

\texttt{Generate GRASP92-like T coefficients for scalar
interactions ?}.

\medskip

appears. The response \texttt{y} or \texttt{Y} causes
GRASP92--like
$T_{rs}(ab)$ coefficients to be generated, whereas the
alternative
\texttt{n} or \texttt{N} yields $t^0_{rs}(ab)$
coefficients.

\medskip

A similar dialog follows for two--electron angular coefficients.
A number of examples illustrate the usage of ANCO in this mode
in section 4 below.

\medskip

The prompt

\medskip

\texttt{Enter a file name for the  anco.vnu  file:}

\medskip

permits the user to specify where the spin--angular coefficients
should be stored.

\begin{figure}
\begin{small}
\begin{verbatim}
 ANCO: Calculation of angular coefficients for symmetry-adapted CSF functions
  from the GRASP92 structure program (Fortran 90 version)
  (C) Copyright by G Gediminas and others, Kassel (2000).

 Enter a file name for the  anco.sum  file:
test.sum
 Enter the name of the configuration symmetry list file:
argon-sd.inp
 Loading configuration symmetry list file ...
  There are  16  relativistic subshells;
  there are  761  relativistic CSFs;
  ... load complete.
 Generate only non-trivial angular coefficients which include
 (at least one) open shells ?
y
 Generate one-electron angular coefficients for scalar interactions ?
y
  Generate GRASP92-like T coefficients for scalar interactions ?
y
 Generate two-electron angular coefficients for scalar interactions ?
y
  Generate GRASP92-like V^k coefficients for scalar interactions ?
y
 Enter a file name for the  anco.vnu file:
test.vnu
\end{verbatim}
\end{small}
{\bf Figure 1:}
\hspace{0.2cm}
{\rm The typical interactive dialog for calculation spin--angular
coefficients.}
\end{figure}

\medskip

The module ANCO needs one input file \texttt{.csl} containing the CSF
list output by GRASP92~\cite{GRASP92}, which is gnerated by the
program GENCSL. 

\medskip

The program creates two output files. The file \texttt{.sum} contains
the summary of the problem. The other file \texttt{.vnu}
contains the
angular momentum coefficients and their characteristics. Its format
for $T_{rs}(ab)$ and $V_{rs}^{k}(abcd)$ coefficients
is the same as that of file \texttt{genmcp.dbg} generated by the
module \texttt{genmcp} of GRASP92~\cite{GRASP92}. The format
for $t^0_{rs}(ab)$ and $v_{rs}^{k}(abcd)$ is very similar
but without the sorting
process which is used in GRASP92.
\subsection{Distribution and installation of the program}

As a new component of the RATIP package \cite{Fritzsche/CFF/Dong:99}
similar to the module RCFP~\cite{rabs_rcfp} ANCO will be distributed
as a \texttt{tar} archive file of the
directory  \texttt{ratip\_anco}. On a UNIX (or compatible)
workstation, the command \texttt{tar -xvf ratip\_anco.tar}
reconstructs the file structure. The directory
\texttt{ratip\_anco} then contains the Fortran 90/95
modules
\texttt{rabs\_anco.f} and \texttt{rabs\_recoupling.f},
the program
\texttt{xanco.f} (the main program for interactive work) as
well as the
makefile \texttt{make-anco}. It also includes a number of examples in
the subdirectory \texttt{test-anco} and a short
\texttt{Read.me}
which
explains further details about the installation. Since the same
file structure is preserved in both cases, the combination of ANCO
with
RATIP is simply achieved by running the command \texttt{cp -r
ratip\_anco/. ratip/.} inside the RATIP root directory; then
\texttt{make -f make-anco} will generate the
executable
\texttt{xanco},
together with the other two components
\texttt{xcesd99}
\cite{Fritzsche/Anton:99} and \texttt{xreos99}
\cite{Fritzsche/CFF/Dong:99} of the RATIP package.
The name of the (Fortran 90/95) compiler and special compiler flags
can be overwritten in the header of the makefile as
necessary. Although ANCO uses six other modules which are part already
of RATIP, no further adaptation of the program is needed. At present,
the ANCO program  has been installed and tested under the Linux and
AIX  operating systems but, owing to the compliance of the Fortran
90/95 standard, no difficulties should arise on any other
platform.

\medskip

The subdirectory \texttt{test-anco} lists a number of examples
which demonstrate the usage of the program.

\section{Timing and verification of \texttt{ANCO}}

Tests and timing studies using the Dirac-Coulomb Hamiltonian only were
performed for the $3s^{2}3p^{6}$ $^{1}S$ 
state of Ar I with the common closed shells $1s^{2}2s^{2}2p^{6}$
for different values of final orbital momentum $J$. The wave
function expansions used were:

\begin{enumerate}

\item 3SD: Single and double excitations from $3s^{2}3p^{6}$ to 
the active set
$ \{ 3s,3p,3d \} $
 contains 14 configuration state functions (CSF) for $J= 0$
and 34 CSF (the maximum) for $J= 2$.

\item 3SDT: Single, double and triple excitations from $3s^{2}3p^{6}$ to
the active set $\{ 3s,3p,3d \} $.
 The maximum number of CSF is 145 for $J= 2$.

\item 4SD: Single and double excitations from $3s^{2}3p^{6}$ to the 
active set
\\$ \{ 3s,3p,3d,4s,4p,4d,4f\} $. 
The maximum number of CSF is 465 for $J= 2$.

\item 4SDT: Single, double and triple excitations from $3s^{2}3p^{6}$
  to the active set\\ $\{ 3s,3p,3d,4s,4p,4d,4f \} $.

\item 5SD; Single and double excitations from $3s^{2}3p^{6}$ to the 
active set
\\$\{ 3s,3p,3d,4s,4p,4d,4f,5s,5p,5d,5f,5g \}$. 

\end{enumerate}

We first considered simple cases with a small number of CSF (3SD, 3SDT,
4SD with $J$ = 0, 1, 2, 3, 4, 5, 6, 7, 8, 9).  Although ANCO
generates the full set of ``pure'' coefficients for both one-- 
and
two--particle operators, the calculation runs from 1.4 -- 2.3 times
faster than an equivalent calculation with \texttt{GRASP92} because of
the lower computational overheads.  Table~\ref{run_time} 
demonstrates
similar enhanced performance for the much larger 4SDT and 5SD 
eamples,
showing the improvement expected for large--scale calculations. 

\medskip

From the results presented in the Table~\ref{run_time} we conclude
that the new program is not much faster in simple cases, but does
better in more complicated cases. The fact that \texttt{ANCO}
calculates approximately twice the number of angular coefficients as
\texttt{GRASP92} reduces the effective cost per coefficient by a
further factor of two. 

\medskip

Although the program is completely new, we have verified that the
results presented agree completely with those obtained from
GRASP92.  We have also verified that the Breit interaction is 
treated
correctly, although no data are presented here.

\begin{table}
\begin{center}
\caption{Timing Comparison for {\texttt{GRASP92}} and
{\texttt{ANCO}} codes.  Times are given in hours, minutes, seconds}
\vspace{5 mm}
\label{run_time}

\begin{tabular}{|r|r|r|r|r|l|l|l|} \hline
&\multicolumn{4}{c|}{} & \multicolumn{2}{c|}{}&  \\ [-0.2cm]
{\bf ASF} & \multicolumn{4}{c|}{\bf Number of}
& \multicolumn{2}{c|}{\bf Running time of}&{\bf Speed} \\ [0.3cm]
 \cline{2-7}
{\bf expan.}& & & & & & & {\bf -up} \\ [-0.2cm]
& {\bf CSF} & {\bf $T_{rs} ^{k} (ab)$ or} & {\bf $V_{rs} ^{k} (abcd)$}
& {\bf $v_{rs} ^{k} (abcd)$}
& {\texttt{GRASP92}}
& {\texttt{ANCO}}
&
\\
&   &  {\bf $t_{rs} ^{k} (ab)$ } &
& &
&   &
\\ [0.3cm] \hline
& & & & & & & \\ [-0.2cm]
4SDT (J= 0)&2~149& 3~606&  756~023& 1~530~086&00:08:11&
00:03:01&2.7 \\
4SDT (J= 1)&5~786&14~017&4~070~156&
 8~188~130&00:59:01&00:15:19&3.9 \\
4SDT (J=
 2)&8~016&21~356&7~018~885&14~077~044&01:42:47&00:26:28&3.9 \\
4SDT (J=
 3)&8~378&21~342&7~634~136&15~290~955&01:53:55&00:30:37&3.7 \\
4SDT (J=
 4)&7~284&15~971&6~111~074&12~260~139&01:33:17&00:23:01&4.1 \\
4SDT (J= 5)&5~349& 9~435&3~810~165&
 7~656~054&00:50:27&00:14:18&3.6 \\
4SDT (J= 6)&3~370& 4~556&1~836~602&
 3~706~544&00:21:52&00:06:40&3.3 \\
4SDT (J= 7)&1~788& 1~789&  693~761&
 1~412~443&00:07:26&00:02:29&3.0 \\
5SD  (J= 0)&  468&   621&   75~192&
 150~455&00:00:32&00:00:17&1.9 \\
5SD  (J= 1)&1~134& 2~324&  395~450&
 792~560&00:03:10&00:01:29&2.1 \\
5SD  (J= 2)&1~609& 3~704&  697~651&
 1~395~839&00:06:27&00:02:44&2.4 \\
5SD  (J= 3)&1~584& 3~441&  721~907&
 1~444~095&00:06:43&00:02:59&2.3 \\
5SD  (J= 4)&1~361& 2~500&  558~223&
 1~117~681&00:05:15&00:02:17&2.3 \\
5SD  (J= 5)&  920& 1~361&  314~909&
 632~306&00:02:30&00:01:22&1.8 \\
5SD  (J= 6)&  559&   644&  141~328&
 284~102&00:01:02&00:00:36&1.7 \\
5SD  (J= 7)&  259&   226&   44~137&
 89~398&00:00:15&00:00:12&1.3 \\
\hline
\end{tabular}
\end{center}
\end{table}

\section{Examples}

To illustrate the use of ANCO in its interactive mode, we studied Ar II.
The program \texttt{genmcp} of GRASP92~\cite{GRASP92} is used
first to
generate the CSF file \texttt{argon-sd.inp} before running
the program
\texttt{xanco}.

\medskip

The first example does a GRASP92--style calculation.  After checking
all triangular conditions for
$X_{Coulomb}^{k}(\alpha \beta \gamma \delta)$ (see 
expression (88)
in~\cite{Grant-a}), it multiplies each pure--two particle coefficient
by the factor $X_{Coulomb}^{k}(\alpha \beta \gamma \delta)$ and prints
all non zero $V_{rs}^{k}(abcd)$ coefficients. With
corrsponding
$T_{rs}(ab)$ coefficients, this generates a total of 11279
coefficients.
The second example, with the same input, calculates a total of
433911 pure
non--trivial spin--angular coefficients at one go, as is more
convenient for large scale calculations.  However, only  11279
coefficients from this set are required in the first example.

\medskip

The Test Run Output displays the \texttt{.sum} files and the
first 10 lines of
\texttt{.vnu} files for both
examples.

\begin{small}

\newpage

\section*{TEST RUN
 OUTPUT}

\begin{small}
\begin{verbatim}
...........................Example 1...........................

>>type example1.sum
 ANCO run at 15:50:39 on Feb 14 2000.

There are  17  electrons in the cloud
 in  761  relativistic CSFs
 based on  16  relativistic subshells.
 Total number of pair is: 289941

Generate only not trivial angular coefficients
 there are 716  Grasp92-like T coefficients
 there are 112078  Grasp92-like Vk coefficients
 the total number of coefficients is 112794

>> type example1.vnu            (first 10 lines of example1.vnu file)
ANCO
761 16 4
V^[( 2)]_[    1,    1] ( 3p , 3p ; 3p , 3p ) =  -1.200000000000E-01
V^[( 1)]_[    1,    1] ( 3p , 3d-; 3d-, 3p ) =   6.666666666667E-02
V^[( 3)]_[    1,    1] ( 3p , 3d-; 3d-, 3p ) =  -2.571428571429E-01
V^[( 1)]_[    2,    1] ( 3p , 3d ; 3d-, 3p ) =  -1.632993161855E-01
V^[( 2)]_[    3,    1] ( 3p , 4s ; 3p , 3d-) =   2.529822128135E-01
V^[( 1)]_[    3,    1] ( 3p , 4s ; 3d-, 3p ) =  -2.108185106779E-01
T^[    6,    1] ( 4d-, 3d-) =    1.000000000000E+00
V^[( 0)]_[    6,    1] ( 1s , 4d-; 1s , 3d-) =   2.000000000000E+00

...........................Example 2...........................

>>type example2.sum
ANCO run at 16:35:02 on Feb 14 2000.

There are  17  electrons in the cloud
 in  761  relativistic CSFs
 based on  16  relativistic subshells.
 Total number of pair is: 289941

Generate only not trivial angular coefficients
 there are 716  pure one-particle angular coefficients
 there are 433195  pure two-particle angular coefficients
 the total number of coefficients is 433911

>> type example2.vnu             (first 10 lines of example2.vnu file)

ANCO
761 16 4
 pure two-particle [( 1)]_[  1,  1] ( 3p , 3p ; 3p , 3p ) =  5.000000000000E-02
 pure two-particle [( 2)]_[  1,  1] ( 3p , 3p ; 3p , 3p ) = -1.500000000000E-01
 pure two-particle [( 3)]_[  1,  1] ( 3p , 3p ; 3p , 3p ) =  5.000000000000E-02
 pure two-particle [( 1)]_[  1,  1] ( 3p , 3d-; 3p , 3d-) =  3.000000000000E-01
 pure two-particle [( 3)]_[  1,  1] ( 3p , 3d-; 3p , 3d-) = -2.000000000000E-01
 pure two-particle [( 0)]_[  1,  1] ( 3p , 3d-; 3d-, 3p ) = -2.500000000000E-01
 pure two-particle [( 1)]_[  1,  1] ( 3p , 3d-; 3d-, 3p ) = -2.500000000000E-01
 pure two-particle [( 2)]_[  1,  1] ( 3p , 3d-; 3d-, 3p ) = -1.500000000000E-01
\end{verbatim}
\end{small}
\medskip

\end{small}
\end{document}